\documentclass[eqsecnum,superscriptaddress,showpacs]{revtex4}
\usepackage{graphics}
\usepackage{amsmath,amssymb}
\usepackage{bm}
\begin{document}
\title{Electric Dipole Moments of Neutron-Odd Nuclei }

\author{Takehisa Fujita}\email{fffujita@phys.cst.nihon-u.ac.jp}
\author{Sachiko Oshima}\email{oshima@phys.cst.nihon-u.ac.jp}
\affiliation{Department of Physics, Faculty of Science and Technology, 
Nihon University, Tokyo, Japan}


\begin{abstract}
The electric dipole moments (EDMs) of neutron-odd nuclei with even protons are 
systematically evaluated. We first derive the relation between the EDM and 
the magnetic moment operators by making use of the core polarization scheme. 
This relation enables us to calculate the EDM of neutron-odd nuclei without 
any free parameters.  From this calculation, one may find the best atomic system 
suitable for future EDM experiments.  

\end{abstract}


\maketitle

\noindent
{\it 1. Introduction-- } 
The discovery of the violation of time reversal invariance in fundamental 
interactions is one of the most important subjects in modern physics. 
For this purpose, the measurement of neutron EDM must be best suited for the 
confirmation due to its simplicity. However, the life time of neutron 
is rather short for accurate measurements, and therefore it is natural that 
the EDM measurement of nucleus should become the next target. Up to now, 
experimentally speaking, the nuclear EDM in neutral atomic systems must be 
the  best for all, but theoretically it is a dangerous attempt since 
the nuclear EDM in neutral atomic systems must be mostly shielded by atomic 
electrons  \cite{q3,q13,muto}. The main physical reason of the shielding should 
come from the basic interaction of the EDM since one starts from the following 
Hamiltonian density due to the Lorentz invariance as well as the gauge invariance
$$ {\cal H}_{edm} 
= - {i\over 2}d^{(i)} \bar \psi^{(i)} \sigma_{\mu \nu} \gamma_5 \psi^{(i)} 
F^{\mu \nu} \eqno{(1)} $$ 
where $ F^{\mu \nu} $ denotes the electromagnetic field strength.  
$d^{(i)}$ is the coupling constant of the EDM interaction for the corresponding 
fermion $\psi^{(i)} $, and here it corresponds to the nucleon field. This Hamiltonian 
density can be reduced to the non-relativistic EDM Hamiltonian 
$$ H_{edm}=-\bm{d}^{(i)} \cdot \bm{E} \eqno{(2)}  $$
where $\bm{d}^{(i)}$ is written as
$$  \bm{d}^{(i)}=d^{(i)} \bm{\sigma}^{(i)} .  \eqno{(3)} $$
If the fermion is neutron, then $d^{(n)}$ is connected to 
the EDM of neutron $d_n$ as $ d_n = d^{(n)} $. 

The main aim of the EDM study is to determine the finite value of $d_n$ by experiments.  
Until now, the observed upper limits of the neutron EDM $ d_n $ are around \cite{q1,harris}
$$ d_n \simeq (0.3\pm 0.5)\times 10^{-25} \  {e\  \rm cm}, \ \ \ 
{\rm Smith\  et\  al.} \eqno{(4)} $$
$$ d_n \simeq (1.9\pm 5.4)\times 10^{-26} \  {e\  \rm cm}, 
\ \ \ {\rm Harris \ et\  al.}  \eqno{(5)}  $$ 
and they are still consistent with zero EDM of neutron. 
At the same time, the experimental efforts to measure the atomic EDM have also been 
performed \cite{w2,w22,q2,q22,q12}. In particular, the EDM of $^{129}{\rm Xe}$ 
have been measured and the observed upper limit of the Xe EDM value is found to 
be \cite{w2,w22}
$$ d_{\rm Xe} \simeq (0.7\pm 3.3 \pm 0.1)\times 10^{-27} \  
{\rm e}\cdot {\rm cm} . \eqno{(6)} $$
In addition, the upper limit of the EDM of Hg atomic system is obtained more accurately 
as \cite{q2,q22}
$$ d_{\rm Hg} \simeq (0.49\pm 1.29 \pm 0.76)\times 10^{-29} \  
{\rm e}\cdot {\rm cm} \eqno{(7)}  $$ 
and one can see that the nuclear EDM measurements in these atomic systems are 
quite precise in comparison with the neutron EDM measurement. 

Last several years, the theoretical investigation of the nuclear EDM in neutral 
atomic systems has been made quite intensively 
\cite{onf,ofa,dks,liu,oshima1,oshima2,itoi}. 
By now, it is clarified that the extraction of the neutron EDM from 
the neutral atomic system is almost impossible due to the electron shielding mechanism 
which is quite a general physical process in all neutral systems. This is somewhat 
unfortunate since the EDMs of the neutral atomic system have been measured much more 
precisely than the ultra-cold neutron experiment by three orders of magnitude, 
but these EDMs may not be related to the neutron EDM because of the shielding mechanism.


However, a recent study indicates that the EDM of ionic systems 
can be directly related to the neutron EDM with some reduction 
factors \cite{oshima1,oshima2,itoi}. In this case, the electron shielding mechanism 
becomes incomplete, and 
thus the EDM of ions can be described by the neutron EDM without having any 
further complications. 

In this report, we discuss the EDM of the ionic systems in which two electrons are 
stripped off. In this case, the reduction factor becomes 
$2/Z$, and it should be noted that, if one makes ions 
where one electron is stripped off, then the reduction factor becomes 
$1/Z$, and so on. In this case, we assume that the atomic state 
should have the spin zero state which does not affect on the spin precession.

\vspace{0.2cm}
\noindent
{\it 2. EDM of Ions in shell model calculations-- } 
In ions with two electrons stripped off, the nuclear EDM can be described 
as \cite{itoi,oshima2}
$$  d_A ={2\over Z} \langle \Psi_0| \sum_{i=1}^A d^{(i)}\sigma_z^{(i)} |\Psi_0 
\rangle \eqno{(8)} $$
where $A$ and $Z$ denote the mass number and proton number of nucleus, respectively. 
$\Psi_0$ denotes the nuclear ground state. This is the nuclear EDM term which is escaped 
from the Schiff theorem. 

The calculations of the nuclear EDM in ions can be reduced to the evaluation of 
the spin matrix elements. The shell model calculations of the spin operators have been 
made in many nuclear processes \cite{verga,rev,suh}. Here, the calculations can be 
carried out by making use of the experimental values of the nuclear magnetic 
moments \cite{engel2}. This is clear since the magnetic moment operator in nucleus 
can be written as
$$ \bm{\mu} = \mu_N \sum_{i=1}^A  (g_s^{(i)} \bm{s}^{(i)}+g_\ell^{(i)} 
\bm{\ell}^{(i)} ) = \mu_N \sum_{i=1}^A  ( (g_s^{(i)}-g_\ell^{(i)}) \bm{s}^{(i)}+
g_\ell^{(i)} \bm{j}^{(i)} ) \eqno{(9)} $$
where $\mu_N$ denotes the nuclear magneton, and hereafter, we set $\mu_N=1$.  
$g_s^{(i)}$ and $g_\ell^{(i)}$ denote the g-factors of nucleon. 
For the neutron odd nucleus with even numbers of protons, we can evaluate 
the magnetic moment $\mu_A$ in terms of the configuration 
mixing method \cite{arima} which is the modification only for the spin operators because  
there is no effect on the $\bm{j}^{(i)}$ operators from the core polarization. 
Thus, the magnetic moment $\mu_A$ can be written as \cite{noya}
$$ \mu_A = \langle s_z \rangle ( g_s^{(n)}+\delta g_s ) , \ \ \ \ \ 
\langle s_z \rangle =  \begin{cases}
 {1\over 2} & {\rm for} 
\ \ \ \ j=\ell+{1\over 2} \\
-{1\over 2}{2\ell-1\over 2\ell +1} &  {\rm for} \ \ \ \ j=\ell-{1\over 2} 
\end{cases}  \eqno{(10)}  $$
where $\delta g_s$ denotes the effective $g-$factor which arises from the perturbation 
calculation and is given as
$$ \delta g_s =-{8\over 3}\alpha_p N_p(g^{(p)}_s -1){2\ell_p(\ell_p+1)\over 2\ell_p+1} 
-{8\over 3}\alpha_n N_n g^{(n)}_s {2\ell_n(\ell_n+1)\over 2\ell_n+1} . \eqno{(11)} $$
Here, $\ell_p$ ($\ell_n$) and $N_p$ ($N_n$) denote the angular momentum and 
the number of the core protons (neutrons) which can contribute 
to the configuration mixing. $\alpha_p$ and $\alpha_n$ are related to the residual 
nucleon-nucleon interaction, and for simplicity, we make them as free parameters so as to 
reproduce the observed magnetic moments of nucleus. From the property 
of the residual interaction, it is safe to assume the following relation between 
$\alpha_p$ and $\alpha_n$
$$ \alpha_p=-3\alpha_n \equiv -3\alpha \eqno{(12)} $$
where $\alpha$ is the only free parameter. The experimental values of 
$g-$factors are given $ g_s^{(p)}=5.585$ and $ g_s^{(n)}=-3.826 $. 
By making use of the value of $\alpha$ determined from the observed nuclear magnetic 
moments, we can evaluate the EDM for the neutron odd nucleus with the mass number $A$ 
$$ d_A = \left({2\over  Z} \right)\langle s_z \rangle \left(d_n -{8\over 3}\alpha_p N_p 
d_p {2\ell_p(\ell_p+1)\over 2\ell_p+1} -{8\over 3}\alpha_n N_n
d_n {2\ell_n(\ell_n+1)\over 2\ell_n+1} \right)  \eqno{(13)} $$
where $d_p$ denotes the proton EDM, and as one sees, this is the parameter free 
calculation. Since the observed magnetic moments can be reproduced by the configuration 
mixing method within 20 \% errors, the evaluation of eq.(13) can be also reliable 
in the same level of accuracy. 

In Table 1, we present the calculated results of the EDM for the neutron odd nucleus 
with even protons for wide ranges of nuclear chart. From this Table, one can hopefully 
find the best possible candidate for the EDM experiment with ions which should have 
the spin zero atomic state. 

\vspace{0.2cm}


\begin{center}
\underline{Table 1} \\
\ \ \\
\begin{tabular}{|c|c|c|c|c|c||c|}
\hline
 \ \ \ Nucleus \ \ \ & \ \ \ \ $A$  \ \ \ \  &  \ \ \ \ $Z$  \ \ \ \  
&  \ \ \ \ $J^P$  \ \ \ \ &   \ \ \ \ \ \ \  $\mu_A^{exp} $   \ \ \ \ \ \ \  
& $\alpha$ &   \qquad \qquad  $ d_A $   \qquad \qquad   \\
\hline
\hline
Be \ & 9  & 4  & ${3\over 2}^- $ & $-1.1778$ & 0.0117 & $0.23 d_n+0.063d_p$  \\
\hline
C \ &13 & 6 & ${1\over 2}^- $   & $0.7024$  & $-0.0016$ & $-0.057 d_n+0.0037d_p$   \\
\hline
O \ & 17 & 8 & ${5\over 2}^+ $   & $-1.89379$ & 0&  $0.125 d_n$   \\
\hline
Mg \ & 25 & 12 & ${5\over 2}^+ $  & $-0.8555$ & 0.0047 & $0.073 d_n+0.030d_p$    \\
\hline
Si \ & 29 & 14 & ${1\over 2}^+ $  & $-0.5553$ & 0.0040 & $0.060 d_n+0.033d_p$    \\
\hline
S \ & 33 & 16 & ${3\over 2}^+ $  & $0.6438$ & 0.0025 &  $-0.034 d_n-0.011d_p$    \\
\hline
Ar \ & 39 & 18 & ${7\over 2}^- $  & $-1.3$ & 0.00046 &  $0.056 d_n+0.015d_p$    \\
\hline
Ca \ & 43 & 20 & ${7\over 2}^- $  & $-1.3176$ & 0.017 &  $0.034 d_n$    \\
\hline
Ni \ & 61 & 28 & ${3\over 2}^- $  & $-0.750$ &0.0018 &   $0.031 d_n+0.014d_p$   \\
\hline
Zn \ & 67 & 30 & ${5\over 2}^- $  & $0.875$ & 0.0090 &  $-0.021 d_n-0.0046d_p$    \\
\hline
Ge \ & 73 & 32 & ${9\over 2}^+ $  & $-0.879$ & 0.0035 &  $0.031 d_n+0.014d_p$   \\
\hline
Sr \ & 87 & 38 & ${9\over 2}^+ $  & $-1.094$ & 0.0038 &  $0.017 d_n+0.0014d_p$   \\
\hline
Zr \ & 91 & 40 & ${5\over 2}^+ $  & $-1.304$ & 0 &   $0.025 d_n$   \\
\hline
Sn \ & 119 & 50 & ${1\over 2}^+ $  & $-1.047$ & 0.00012 &  $0.020 d_n+0.0076d_p$  \\
\hline
Te \ & 125 & 52 & ${1\over 2}^+ $  & $-0.889$ & 0.0010 &  $0.016 d_n+0.0057d_p$   \\
\hline
Xe \ & 129 & 54 & ${1\over 2}^+ $  & $-0.778$ & 0.0012 &  $0.015 d_n+0.0065d_p$   \\
\hline
Ba \ & 135 & 56 & ${3\over 2}^+ $  & $0.838$ & 0.00052 &  $-0.0098 d_n-0.0016d_p$   \\
\hline
Nd \ & 143 & 60 & ${7\over 2}^- $  & $-1.066$ & 0.0023 &  $0.010 d_n+0.0007d_p$   \\
\hline
Hg \ & 201 & 80 & ${3\over 2}^- $  & $-0.560$ & 0.00092 &  $0.0097 d_n+0.0051d_p$   \\
\hline
\end{tabular} 


\vspace{0.5cm}
\begin{minipage}{13cm}

Table 1 shows the calculated values of the nuclear EDM $d_A$ for odd-neutron nuclei 
in the $A^{++}$ ionic states where two electrons are stripped off. 
The observed values of the nuclear magnetic moments and the calculated values 
of $\alpha$ are also shown \cite{iso}. 

\end{minipage}
\end{center}

\vspace{0.5cm}
\noindent
{\it 3. Discussions and Remarks-- } 
By now, we have learned that the nuclear EDM in neutral atomic systems must be mostly 
shielded by atomic electrons. Intuitively, electrons always block the penetration of 
the electric field inside nucleus, and unless one excites atomic electron states, one 
cannot find the penetration of the electric field inside nucleus. 
But, still, from the excitation of electrons, one can find only a very small 
flux of electric field inside nucleus. 

This shielding mechanism can be well explained in the following fashion. 
In the Schiff screening, the nucleon EDM operator 
($ \sum_i^A \bm{d}^i \cdot \bm{E}_{ext}  $) cannot be seen because electrons 
trapped in the Coulomb filed of $A_0(r_j)$  always react as
$$ \sum_{i=1}^A  \sum_n  \langle  \psi_0 |\sum_{j=1}^Z{ \bm{d}^i} \cdot  
\bm{\nabla} A_0(r_j)  |\psi_n \rangle {1 \over E_0 - E_n} 
 \langle \psi_n | \sum_{k=1}^Z e \bm{r}_k 
\cdot \bm{E}_{ext} |\psi_0 \rangle + h.c.  \eqno{(14)}  $$
which can be just rewritten as ($ -\sum_i^A \bm{d}^i \cdot \bm{E}_{ext}  $), and 
this cancels out completely the original nucleon EDM operator. 
Here, $\psi_n $ denotes the atomic wave function.  
Now, one can easily see that this Schiff theorem cannot hold for the nuclear EDM 
in hydrogen-like atom with $Z$ different from unity \cite{itoi}. This is clear 
since the proof is based on the Coulomb field for electron, but the electric field 
on to the nuclear charge is produced by electron, and this is different from 
the $A_0(r)$ potential. Therefore, unless the charge $Z$ is equal to unity, 
the nuclear EDM survives while the electron EDM always vanishes to zero due to 
the Schiff theorem. 

In addition, the cancellation due to the electron screening takes place 
for nuclear dipole operators ($ \sum_i^Z e \bm{R}_i \cdot \bm{E}_{ext} $) 
since the following higher order effects 
$$ \sum_{i=1}^Z \sum_n \langle \psi_0 |\sum_{j=1}^Z  {e^2(\bm{r}_j \cdot \bm{R}_i  )
\over r_j^3 } |\psi_n \rangle {1 \over E_0 - E_n} 
 \langle \psi_n| \sum_{k=1}^Z e \bm{r}_k 
\cdot \bm{E}_{ext} |\Psi_0 \psi_0 \rangle + h.c.  \eqno{(15)} $$ 
can be reduced to ($- \sum_i^Z e \bm{R}_i \cdot \bm{E}_{ext}  $) which  
cancels out completely the original nuclear dipole operator \cite{ofa,dks,liu,oshima1}.  
Therefore, as long as we consider the nuclear EDM operator, we cannot observe 
the EDM of nucleon from the neutral atomic system. This proof is done at the level 
of operators for the nuclear variables, and therefore, it is impossible to extract 
the nuclear EDM from the neutral atomic systems.  
Here, we should note that the finite nuclear EDM in the neutral systems can be obtained 
only through the atomic excitation in the intermediate states where the nuclear state 
is taken to be in the ground state. In this case, one 
obtains the nuclear EDM \cite{ofa,oshima1}
$$ d_A =-\sum_n {2e \over{E_n-E_0 }} \langle \psi_0 |  
\sum_{i=1}^Z\left[\bm{ r}_i \cdot  \sum_{j=1}^A \bm{ d}^j 
\left({5\over 2}-{15\over 2}\cos^2 \theta_{ji}\right)
R_j^2  \right]{e\over{r_i^5}}  
|\psi_n  \rangle \langle \psi_n |\sum_{i=1}^Z z_i |\psi_0  \rangle  
\simeq 2.4 \times 10^{-6} d_n . 
\eqno{(16)} $$
This suppression factor 
of $2.4 \times 10^{-6}$  is too small to carry out the EDM measurement so as to 
compete with the direct neutron measurement. In addition, the above estimation is 
made by using optimistic values for the parameters in the calculation. Also, one can 
easily see that the main reason of this suppression comes from the ratio between 
atomic radius $a_0$ and nuclear radius $R_0$, and the factor is proportional to 
$ \sim (R_0/a_0)^2 $ which is quite small. This suppression factor is indeed 
confirmed by the new calculation of Yoshinaga et al. \cite{yoshi} who carried out 
quite elaborate shell model calculations in Xe nucleus. Therefore, practically, 
there is no chance to observe the nuclear EDM in the neutral atomic system. 

In order to overcome this difficulty, we should go on to the measurement of 
nuclear EDM in ionic systems as we show here. Therefore, one should strip a few 
electrons from atoms, and then one can measure the nuclear EDM which is directly 
related to the nucleon EDM. A question is, of course, as to how people can measure 
the nuclear EDM of ions which are certainly affected by the electric field 
$ \bm{E}_{ext}$. This should be solved by experimentalists, and indeed Orlov et al. 
presented an interesting proposal to measure the EDM of ions in the storage ring 
experiments \cite{sto}. This type of experiments should become very important 
in future, but at this moment, we cannot claim anything further, and should wait 
for any EDM measurements from the storage ring experiments or any other EDM 
experiments of ionic systems.

\vspace{0.8cm}

\noindent
S.O. would like to thank the Japan Society for the Promotion of Science 
for financial support.


\end{document}